\documentclass{statsoc}

\usepackage{etoolbox}
\makeatletter
\patchcmd{\@makecaption}
  {\parbox}
  {\advance\@tempdima-\fontdimen2} 
  {}{}
\makeatother 

\usepackage{amsmath,amsfonts,color,enumerate,alltt,mathrsfs}
\usepackage{tabularx,booktabs}
\usepackage{multirow}
\usepackage{natbib}
\usepackage{pstricks,pst-all}
\usepackage[left=30mm,right=30mm,top=30mm,bottom=30mm]{geometry}

\newcommand{\vect}{\boldsymbol}
\newcommand{\matr}{}
\newcommand{\E}{\mathrm{E}}
\newcommand{\Cor}{\mathrm{Cor}}
\newcommand{\Var}{\mathrm{Var}}
\newcommand{\logit}{\mathrm{logit}}

\newcolumntype{Y}{>{\centering\arraybackslash}X}

\title{Identifying the effect of public holidays on daily demand for gas}
\author[Sarah~.E.~Heaps {\it et al.}]{Sarah~E.~Heaps}
\address{Newcastle University,
Newcastle upon Tyne,
UK.}
\email{sarah.heaps@ncl.ac.uk}
\author[Sarah~.E.~Heaps {\it et al.}]{Malcolm~Farrow}
\address{Newcastle University,
Newcastle upon Tyne,
UK.}
\author[Sarah~.E.~Heaps {\it et al.}]{Kevin~Wilson}
\address{Newcastle University,
Newcastle upon Tyne,
UK.}

\usepackage{xr}
\begin{document}


\begin{abstract}
To reduce operational costs, gas distribution networks require accurate forecasts of the demand for gas. Amongst domestic and commercial customers, demand relates primarily to the weather and patterns of life and work. Public holidays have a pronounced effect which often spreads into neighbouring days. We call this spread the ``proximity effect''. Traditionally, the days over which the proximity effect is felt are pre-specified in fixed windows around each holiday, allowing no uncertainty in their identification. We are motivated by an application to modelling daily gas demand in two large British regions. We introduce a novel model which does not fix the days on which the proximity effect is felt. Our approach uses a four-state, non-homogeneous hidden Markov model, with cyclic dynamics, where the classification of days as public holidays is observed, but the assignment of days as ``pre-holiday'', ``post-holiday'' or ``normal'' is unknown. The number of days to the preceding and succeeding holidays guide transitions between states. We apply Bayesian inference and illustrate the benefit of our modelling approach. A preliminary version of the model is now being used by one of the UK's regional distribution networks.
\end{abstract}


\section{\label{sec:intro}Introduction}

The energy sector in the UK is changing. In order to comply with the UK Climate Change Act 2008, greenhouse gas emissions must be reduced to 80\% of their 1990 levels by the year 2050 \citep[][]{HMP08}. Pragmatically,  these targets must be met in a manner which does not compromise the provision of affordable prices to consumers or the competitiveness of UK industry. At present, natural gas provides a comparatively low cost source of energy which, unlike some renewables such as wind power, can also offer reliability in supply and storability \citep[][]{CM12}. In the future, decarbonisation of the gas network is likely to involve switching from natural gas to carbon-neutral bio-methane or converting the gas distribution network to transport hydrogen rather than natural gas \citep[][]{DM13}. Therefore both now and over the years to come, gas has a vital role to play in the energy mix and it is more important than ever that the gas distribution network operates as efficiently as possible.

National Grid is the sole owner and operator of the gas transmission infrastructure in the UK. Gas in the national transmission system leaves the network at high pressure at 49 points across the country. After being odorised for safety it is transported, ultimately at lower pressure, through eight regional distribution networks to individual customers. National Grid works closely with the regional distribution networks to ensure that the local supply of gas meets the demand at all times.


Demand forecasts, over a range of horizons, are required by both National Grid and the distribution companies for reasons including safety and security of supply and investment and operational planning \citep{NG16}. The method used by distribution companies involves forecasts $\hat{A}_{u,j}$ for the annual demand $A_{u,j}$ in year $u$ from a subset $j$ of customers, $j=1, \ldots, J$. These are produced outside the company, taking into account economic and other external factors. Forecasts for demand $\tilde{Y}_{t,j}$ for subset $j$ on day $t$ are then rescaled to match this annual forecast; see \citet{NG16}. Thus what is required is inference about a set of scale factors $ \exp\{k_{t,j}\}$ for the days of the year so that the forecast mean of $\tilde{Y}_{t,j}$, for example, will be $ {\rm E}[\exp\{k_{y,j}\}]\hat{A}_{u,j}$, since the scale factors may reasonably be treated as independent of the annual demand level.

In this paper we describe a model for demand using the transformation $Y_{t,j}=\ln (\tilde{Y}_{t,j}) = c_{u(t),j} + k_{t,j} + e_{t,j}$ where $u(t)$ is the year in which day $t$ falls. Here $e_{t,j}$ represents random fluctuation and the value of the additive constant $c_{u(t),j}$ can be adjusted in forecasts to match $\hat{A}_{u(t),j}$, without affecting the log scale factors $k_{t,j}$. It is these factors, incorporating weather, seasonal and calendar effects, about which inference is required.

There is a large body of work in the scientific literature concerned with modelling and forecasting the demand for natural gas; see, for example, \citet{Sol12} for an extensive review. In broad terms, consumers of gas can be divided into three types: residential, commercial and industrial. We focus on the first two of these groups in this paper. Although economic factors like gas prices and national income are important drivers of gas consumption by big industrial users, their effects are generally less important for residential and commercial customers for whom gas is predominantly used for heating (including water) and cooking. As a result, the demand by these groups is strongly related to the weather and patterns of life and work. Models for residential and commercial gas consumption therefore generally allow for \emph{weather-related predictors} and \emph{seasonal and calendar effects}. Sometimes models also incorporate interactions between the two to allow the effect of the weather and, in particular, temperature to vary with periodic changes in fixed or timed heating schedules. After modelling the weather, seasonal and calendar effects, it is common for remaining errors to exhibit considerable autocorrelation. This is generally modelled directly, for instance by assuming autoregressive-moving-average (ARMA) models for the residuals \citep[][]{Lyn84,AP16}.

\emph{Weather-related predictors} in models for gas consumption generally include primitive variables, like temperature and wind speed, or derived variables, constructed to have a simple relationship with demand. In the literature, considerable attention has been devoted to modelling the non-linear effect of temperature which manifests at low and high temperatures when, for example, many customers switch their heating on or off. In some cases this is achieved by fitting non-linear regression models \citep[e.g. see][]{BKM09,GS18} whilst in others, the effect is captured through bespoke, derived variables such as ``heating-degree days'' \citep[e.g.][]{AP16}. The latter measure the temperature difference from some fixed threshold, but adopt zero values when the temperature exceeds that threshold, so that the temperature has no effect when heating is not needed.

\emph{Seasonal and calendar effects} often include a smoothed effect for the day of the year \citep[e.g.][]{BKM09,GS18} and fixed effects to represent the day of the week. These are needed because gas demand follows a weekly cycle, with clear differences between weekdays and weekends \citep[][]{Lyn84}. Similarly, because demand is affected markedly by public holidays, models usually include a holiday factor \citep[][]{BKM15} or treat public holidays like days of the weekend \citep[e.g.][]{CS14}. In addition to effects on the days of the holidays themselves, models for energy consumption sometimes include a protracted effect which extends into neighbouring days. This allows for the influence of changes in cooking and travel arrangements, and for the commercial and industrial slow-down that typically occurs around holidays. For example, in their model for gas consumption in the Czech Republic, \citet{BKM09} include Christmas and Easter effects which apply over a fixed range of consecutive days around Christmas Day and Good Friday. Taking a different approach, \citet{BMP10} define a ``daytype'' factor with five levels corresponding to different classifications of the current, previous and following days as working days or otherwise. This allows public holidays and neighbouring days to have different effects depending on where they fall in the week. In their models for electricity consumption, \citet{HWM06} and \citet{PMV02} allow demand to differ on \emph{proximity days} within a short fixed window around each holiday. 

Some public holidays, particularly those around Christmas and often Easter, occur during periods where the demand for gas is at its highest. At such times, it is essential to have accurate forecasts and a correct quantification of uncertainty. Yet, in terms of allowing for a \emph{proximity effect}, an aspect common to all of the approaches above is that the dates of the proximity days are fixed, often arbitrarily. 

In this paper, we are motivated by an application involving daily gas consumption in the two large geographical areas in Northern England served by the regional operator Northern Gas Networks (NGN). We propose a novel model for daily gas demand which, to our knowledge, is the first approach which does not fix the days on which the proximity effect of a public holiday is felt. The results of a preliminary version of this model are already being used by the company in its annual medium-term forecasting exercise. Our approach is based on a four-state, non-homogeneous hidden Markov model (NHMM) with cyclic dynamics. In this model the classification of days as public holidays is observed, but the assignment of days as ``pre-holiday'', ``post-holiday'' or ``normal'' is unknown, but guided by the number of days to the preceding and succeeding public holidays. We allow for auto-- and cross--correlation in the bivariate time-series by modelling the logarithm of gas demand in the two regions, conditional on the states, using a bivariate autoregression of order one, with a symmetric autoregressive coefficient matrix, that allows stationarity to be imposed through simple constraints on the parameter space. Taking a Bayesian approach to inference, we use a hierarchical structure to encapsulate structural prior information about similarities between the two regions. In our application, we illustrate the benefit of allowing for a proximity effect of unknown magnitude and duration.  

The remainder of this paper is structured as follows. The data for our analysis are described in Section \ref{sec:data}. In Section \ref{sec:model} we propose our model and describe it in detail. Section \ref{sec:prior} gives the structure of our prior distribution and Section \ref{sec:post_inf} discusses computation of the posterior distribution. We give the results of applying our model and inferential procedures in Section \ref{sec:appl} and draw some conclusions in Section \ref{sec:discuss}.

\section{\label{sec:data}The data}

NGN, one of the eight regional distribution networks in the UK, is responsible for gas distribution to 2.7 million homes and businesses across a $25000 \mathrm{km}^2$ region in the North East of England, Northern Cumbria and Yorkshire.

In the UK, medium and long term demand forecasts are produced for each of thirteen local distribution zones (LDZs). NGN is responsible for generating the demand forecasts for the neighbouring Northern (NO) and North East (NE) LDZs. The former covers Northern Cumbria and the North East of England and the latter encompasses much of Yorkshire. Within each LDZ, every point at which gas is taken from the network by supply pipe to a consumer is categorised according to its \emph{load band}, which is based on how much gas it uses. Depending on whether or not meters are read daily, load bands are classified as daily metered (DM) or non-daily metered (NDM). DM load bands comprise large industrial premises which typically have the highest demand for gas. NDM load bands are sub-divided into four categories, roughly containing domestic, commercial and small and medium sized industrial customers. In this paper we focus on load bands 1 to 3, the definitions of which are shown in Table~\ref{tab:lb}.

\begin{table}
\caption{\label{tab:lb}Classification of NDM load bands 1 to 3.}
\centering
\begin{tabular}{cll} 
\toprule
\multicolumn{2}{c}{Load Band} &Example \\
Index & MWh / year            &\\
\midrule 
1     & 0-73 MWh              & A single house \\
2     & 73-732 MWh            & A large block of flats or commercial premises \\
3     & 732-5860 MWh          & Small industrial premises \\ 
\bottomrule
\end{tabular}
\end{table}

For both LDZs and each of the three NDM load bands, daily gas consumption data are available for the nine year period from 1st January 2008 to 18th February 2017. The accompanying weather data take the form of a derived variable called the \emph{composite weather variable} (CWV), defined by National Grid, which takes a single, daily value for each LDZ. The CWV is based primarily on air temperature. However, in order to strengthen the linear association between this temperature-based variable and gas consumption, its construction accommodates adjustments at high and low temperatures and allowance for the effects of other weather variables, such as wind speed. Further details can be found in \citet{NG16}.

\section{\label{sec:model}A non-homogeneous hidden Markov model (NHMM)}

Denote by $\tilde{Y}_{t,j}$ the gas demand in tenths of a gigawatt-hour (GWh) pertaining to a given NDM load band on day $t$ in region $j$, where $j=1$ and $j=2$ correspond to the Northern and North East LDZs, respectively. Rather than modelling the raw demand, we instead work on a log scale, defining $Y_{t,j} = \ln ( \tilde{Y}_{t,j} )$ and $\vect{Y}_{t}=(Y_{t,1},Y_{t,2})'$ for $t=1,\ldots,T$. This helps to make the variance more stable across seasons and gives fixed effects in our additive model a multiplicative effect on the original scale. The CWV on day $t$ is denoted by $\vect{w}_t = (w_{t,1}, w_{t,2})'$. We introduce further explanatory variables $n_t \in \mathbb{N}_0=\{ 0, 1, 2, \ldots \}$ and $p_t \in \mathbb{N}_0$ which indicate the number of days to the next and since the previous public holiday, respectively. Clearly day $t$ is a public holiday, known colloquially in England as a \emph{bank holiday}, if and only if $n_t=p_t=0$. Finally, we introduce the categorical explanatory variable $r_t \in \{1, 2, 3 \}$ which indicates the ``type'' of the nearest public holiday according to the calendar season in which it falls. The labelling is given in Table~\ref{tab:bank_hols}. Note that if a holiday, such as Christmas Day, falls at the weekend, it is a neighbouring weekday that will be observed as the corresponding bank holiday.

\begin{table}
\caption{\label{tab:bank_hols}Categorisation of UK public holidays. If day $t$ is equidistant between two holidays of different types, the type $r_t$ of the later holiday is used.}
\centering
\begin{tabularx}{0.9\textwidth}{clX} 
\toprule
Type & Name & Days \\ 
\midrule
1 & ``Easter''    & Good Friday and Easter Monday \\
2 & ``Other''     & May Day, Spring bank holiday, Summer bank holiday and one-off holidays such as the Queen's Jubilee \\ 
3 & ``Christmas'' & Christmas Day, Boxing Day and New Year's Day \\
\bottomrule
\end{tabularx}
\end{table}

As discussed in Section~\ref{sec:intro}, public holidays often have a protracted effect on gas consumption, which extends into neighbouring days. The simplest way of modelling this effect would be to classify  days within a fixed window around each public holiday deterministically as proximity days.  However, this approach is inflexible, relying on an arbitrary decision as to the existence, size and position of the windows. Instead, we allow the number of proximity days on either side of each public holiday to be unknown. This is achieved by introducing a discrete-valued stochastic process $\{ S_t: t=0,1,\ldots,T \}$ with $S_t \in \mathcal{S}_s = \{1, 2, 3, 4\}$ in which state 2 can only apply to public holidays and is observable, whilst states 1, 3, and 4 are ``hidden'' and cannot be observed. Of these, states 1 and 3 allow for ``pre-holiday'' and ``post-holiday'' proximity days, whilst state 4 is a baseline for ``normal'' days. To support this we define a distribution over the states so that state 1 can only be entered from state 4 and left via state 2, whilst state 3 can only be entered from state 2.

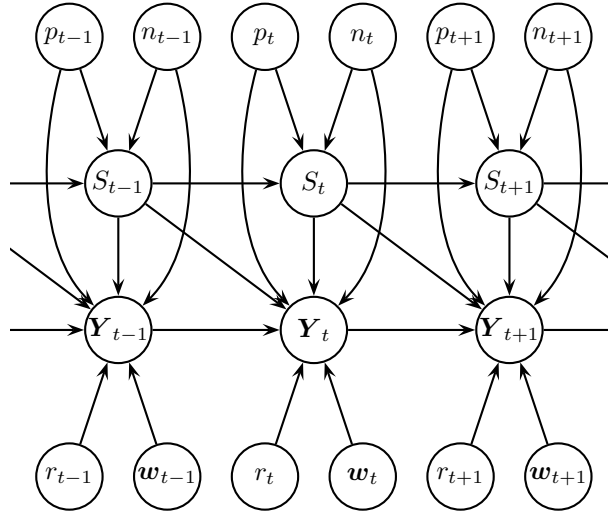
\begin{figure}[t]
\centering
\psset{unit=0.65mm}
\begin{pspicture*}(20,-30)(140,85)
      \scalebox{1}{
        \Cnode[radius=7](30,70){P2}\rput(30,70){\small{$p_{t-1}$}}
        \Cnode[radius=7](70,70){P3}\rput(70,70){\small{$p_{t}$}}
        \Cnode[radius=7](110,70){P4}\rput(110,70){\small{$p_{t+1}$}}
        \Cnode[radius=7](50,70){N2}\rput(50,70){\small{$n_{t-1}$}}
        \Cnode[radius=7](90,70){N3}\rput(90,70){\small{$n_{t}$}}
        \Cnode[radius=7](130,70){N4}\rput(130,70){\small{$n_{t+1}$}}

        \Cnode[radius=7](0,40){S1}\rput(0,40){\small{$S_{t-2}$}}
        \Cnode[radius=7](40,40){S2}\rput(40,40){\small{$S_{t-1}$}}
        \Cnode[radius=7](80,40){S3}\rput(80,40){\small{$S_{t}$}}
        \Cnode[radius=7](120,40){S4}\rput(120,40){\small{$S_{t+1}$}}
        \Cnode[radius=7](160,40){S5}\rput(160,40){\small{$S_{t+2}$}}
        
        \Cnode[radius=7](0,10){Y1}\rput(0,10){\small{$\vect{Y}_{t-2}$}}
        \Cnode[radius=7](40,10){Y2}\rput(40,10){\small{$\vect{Y}_{t-1}$}}
        \Cnode[radius=7](80,10){Y3}\rput(80,10){\small{$\vect{Y}_{t}$}}
        \Cnode[radius=7](120,10){Y4}\rput(120,10){\small{$\vect{Y}_{t+1}$}}
        \Cnode[radius=7](160,10){Y5}\rput(160,10){\small{$\vect{Y}_{t+2}$}}

        \Cnode[radius=7](50,-20){W2}\rput(50,-20){\small{$\vect{w}_{t-1}$}}
        \Cnode[radius=7](90,-20){W3}\rput(90,-20){\small{$\vect{w}_{t}$}}
        \Cnode[radius=7](130,-20){W4}\rput(130,-20){\small{$\vect{w}_{t+1}$}}
        \Cnode[radius=7](30,-20){R2}\rput(30,-20){\small{$r_{t-1}$}}
        \Cnode[radius=7](70,-20){R3}\rput(70,-20){\small{$r_{t}$}}
        \Cnode[radius=7](110,-20){R4}\rput(110,-20){\small{$r_{t+1}$}}
       
        \ncline[arrowscale=1.5]{->}{S1}{S2}
        \ncline[arrowscale=1.5]{->}{S2}{S3}
        \ncline[arrowscale=1.5]{->}{S3}{S4}
        \ncline[arrowscale=1.5]{->}{S4}{S5}
        
        \ncline[arrowscale=1.5]{->}{Y1}{Y2}
        \ncline[arrowscale=1.5]{->}{Y2}{Y3}
        \ncline[arrowscale=1.5]{->}{Y3}{Y4}
        \ncline[arrowscale=1.5]{->}{Y4}{Y5}
        
        \ncline[arrowscale=1.5]{->}{S1}{Y1}
        \ncline[arrowscale=1.5]{->}{S2}{Y2}
        \ncline[arrowscale=1.5]{->}{S3}{Y3}
        \ncline[arrowscale=1.5]{->}{S4}{Y4}
        \ncline[arrowscale=1.5]{->}{S5}{Y5}
        
        \ncline[arrowscale=1.5]{->}{S1}{Y2}
        \ncline[arrowscale=1.5]{->}{S2}{Y3}
        \ncline[arrowscale=1.5]{->}{S3}{Y4}
        \ncline[arrowscale=1.5]{->}{S4}{Y5}
        
        \ncline[arrowscale=1.5]{->}{N2}{S2}
        \ncline[arrowscale=1.5]{->}{P2}{S2}
        \ncline[arrowscale=1.5]{->}{N3}{S3}
        \ncline[arrowscale=1.5]{->}{P3}{S3}
        \ncline[arrowscale=1.5]{->}{N4}{S4}
        \ncline[arrowscale=1.5]{->}{P4}{S4}

        \ncline[arrowscale=1.5]{->}{W2}{Y2}
        \ncline[arrowscale=1.5]{->}{W3}{Y3}
        \ncline[arrowscale=1.5]{->}{W4}{Y4}
        \ncline[arrowscale=1.5]{->}{R2}{Y2}
        \ncline[arrowscale=1.5]{->}{R3}{Y3}
        \ncline[arrowscale=1.5]{->}{R4}{Y4}

        \nccurve[arrowscale=1.5,angleA=-75,angleB=45]{->}{N2}{Y2}
        \nccurve[arrowscale=1.5,angleA=-105,angleB=135]{->}{P2}{Y2}
        \nccurve[arrowscale=1.5,angleA=-75,angleB=45]{->}{N3}{Y3}
        \nccurve[arrowscale=1.5,angleA=-105,angleB=135]{->}{P3}{Y3}
        \nccurve[arrowscale=1.5,angleA=-75,angleB=45]{->}{N4}{Y4}
        \nccurve[arrowscale=1.5,angleA=-105,angleB=135]{->}{P4}{Y4}
      }
\end{pspicture*}
\caption{\label{fig:DAG}DAG illustrating the factorisation of the joint density of the observations and the hidden states conditional on the time series of explanatory variables $(n_t, p_t, r_t, \vect{w}_t)$, for $t=1,2,\ldots$. In a DAG, random variables are represented by nodes. These are connected by directed arrows which indicate the order of conditioning when factorising their joint probability density.}
\end{figure}

The directed acyclic graph (DAG) in Figure~\ref{fig:DAG} illustrates the assumptions of conditional independence in our hidden Markov model. A non-homogeneous Markov chain for the states will be described in Section~\ref{subsec:states}, whilst the model for gas demand, conditional on the states, will be discussed in Section~\ref{subsec:demand}.

\subsection{\label{subsec:states}Modelling the states}
The challenge in modelling the joint distribution of the states arises because state 2, representing public holidays, is observable and the dates of public holidays are always known in advance. In the hidden Markov model (HMM) framework,  we can build this information into our prior distribution for the states $S_t$ in two ways. The more direct approach would be for the states to evolve according to a non-homogeneous Markov chain with transition probabilities that vary over time. Transition into state 2 can occur with certainty if day $t$ is a public holiday, and transition into the pre-holiday state (state 1) can become more likely in the days leading to a holiday. Alternatively we can specify the joint prior distribution for the states by updating an ``initial'' prior, representing a homogeneous Markov chain, by conditioning on the observation that $S_t=2$ if day $t$ is a public holiday or $S_t \ne 2$ otherwise, for $t=0,\ldots,T$. Of course, the resulting process is no longer Markovian. 

We choose to adopt the more direct approach, based on a non-homogeneous Markov chain, and model the transition probabilities as functions of the number of days to the next ($n_t$), and since the previous ($p_t$), public holiday. This approach allows a flexible, non-geometric distribution for the pre- and post-holiday state sojourn times which could, for example, rapidly decay after two days, in keeping with views from the literature (see Section~\ref{sec:intro}) and the expert judgement of engineers at NGN (see Section~\ref{subsubsec:states_other}). Moreover, use of covariates $n_t$ and $p_t$ in the non-homogeneous model allows straightforward experimentation with the joint prior induced for the states, which aids in constructing a distribution with a sensible concentration of prior mass (see Section~\ref{subsec:prior_spec}).


\subsubsection{Public holidays}
Suppose that the states $S_t$ follow a first order, non-homogeneous Markov chain with 
\begin{equation*}
\Pr(S_t = k | S_{t-1} = j, n_t, p_t) = \lambda_{j,k}(n_t, p_t), \qquad (j, k) \in \mathcal{S}_s^2
\end{equation*}
for $t=1, 2, \ldots, T$, and initial distribution
\begin{equation*}
\Pr(S_0 = k | n_0, p_0) = \ell_k(n_0, p_0), \qquad k \in \mathcal{S}_s.
\end{equation*}
Since state 2 is observable, it follows that, if day 0 is a public holiday, then $n_0 = p_0 = 0$ and $\ell_{2}(0, 0) = 1$. Similarly, for any $j \in \mathcal{S}_s$, if day $t$ is a public holiday, then $n_t = p_t = 0$ and $\lambda_{j,2}(0, 0) = 1$. 

\subsubsection{\label{subsubsec:states_other}Other days}
In the remainder of this section we consider the initial distribution $\Pr(S_0 = k | n_0, p_0)$ and transition probabilities $\Pr(S_t = k | S_{t-1} = j, n_t, p_t)$ characterising the state process if day $t$ is not a public holiday. In such cases $(n_t, p_t) \in \mathbb{N}_0^2 \setminus \{ (0, 0) \}$ and state 2 cannot be assigned. Dropping the $t$ subscript for brevity, it follows that $\ell_{2}(n, p) = 0$ and $\lambda_{j,2}(n, p) = 0$ for all $j \in \mathcal{S}_s$.  To complete our initial distribution we assign $\ell_{k}(n, p) = 1/3$ for $k \in \mathcal{S}_s \setminus \{2\}$. Thereafter, we impose cyclic dynamics on the evolution of the state by assuming that all of the remaining transition probabilities $\lambda_{j,k}(n, p)$, for $j \in \mathcal{S}_s$ and $k \in \mathcal{S}_s \setminus \{2\}$, are zero except the transitions from the normal state to the pre-holiday state $\lambda_{4,1}(n, p)$; transitions from the holiday state to the post-holiday or normal states, $\lambda_{2,3}(n, p)$ and $\lambda_{2,4}(n, p)$; and the self-transitions $\lambda_{1,1}(n, p)$, $\lambda_{3,3}(n, p)$ and $\lambda_{4,4}(n, p)$. Necessarily this implies $\lambda_{2,4}(n, p) = 1 - \lambda_{2,3}(n, p)$, $\lambda_{3,3}(n, p) = 1 - \lambda_{3,4}(n, p)$ and $\lambda_{1,1}(n, p) = 1$, leaving us to define $\lambda_{2,3}(n, p)$, $\lambda_{3,4}(n, p)$ and $\lambda_{4,1}(n, p)$.  It follows from the formulation of our model that an uninterrupted spell of proximity days between two public holidays will be deemed post-, rather than pre-, holiday days. This is to avoid difficulties in identifying a time of transition from the post- to the pre-holiday state.


Expert judgement from engineers at NGN suggested that a proximity effect, if it was felt, was likely to last for one or two days on either side of a public holiday. Longer periods of protracted holiday behaviour were thought to be very unlikely. If a pair of public holidays was separated by a small number of ordinary days, transitions into the proximity state, and transitions sustaining it, were expected to be more likely. In order to build the first of these ideas into our model, we allow the probability of transition into the pre-holiday state, $\lambda_{4,1}(n, p)$, to decay quickly to zero as $n$ gets large by modelling the logit of the transition probability as
\begin{equation}\label{eq:lambda_41}
\logit \{ \lambda_{4,1}(n, p) \} = \nu_{4,1,1} + \nu_{4,1,2}\frac{(n - 1)^{1/2}}{10}.
\end{equation}
We anticipate $\nu_{4,1,2} < 0$. Here the offset allows the parameter $\nu_{4,1,1}$ to be interpreted as the logit probability of transition into the pre-holiday state when the following day is a holiday. Taking the square root of $(n-1)$ helps to prevent the prior variance of $\logit \{ \lambda_{4,1}(n, p) \}$ from becoming too large as $n$ grows. In turn, this prevents an implausible $u$-shaped prior for any feasible $n$. Given the values taken by $n$, the division of $(n - 1)^{1/2}$ by 10 allows the coefficient $\nu_{4,1,2}$ to be interpreted on a scale that is comparable with that of a binary predictor, which will be introduced when modelling the other two transition probabilities, $\lambda_{3,4}(n, p)$ and $\lambda_{2,3}(n, p)$. 

For the probability of transition out of the post-holiday state, one could assume $\lambda_{3,4}(n, p) = \lambda_{3,4}$ for all $(n, p) \in \mathbb{N}_0^2 \setminus \{(0, 0)\}$, and appeal to the  property of (homogeneous) first-order Markov chains that the sojourn time in state $3$ would have a geometric distribution with mean $\lambda_{3,4}^{-1}$. However, a geometric distribution of the sojourn time in the post-holiday state, having only one parameter, is not sufficiently flexible to be reconciled with the expert judgement of NGN engineers. We therefore use the information encoded in the explanatory variable $p$, the number of days since the previous holiday, and model the logit of the transition probability as
\begin{equation}\label{eq:lambda_34}
\logit \{ \lambda_{3,4}(n, p) \} = \nu_{3,4,1} + \nu_{3,4,2}\frac{(p - 2)^{1/2}}{10} + \nu_{3,4,3} \mathbb{I}(n = 1)
\end{equation}
in which $\mathbb{I}(n = k)$ takes the value 1 if $n = k$ and 0 otherwise. Clearly the transition probability $\lambda_{3,4}(n, p)$ will approach 1 as $p$ becomes large if $\nu_{3,4,2} > 0$. The offset in the square root term allows the parameter $\nu_{3,4,1}$ to be interpreted as the logit probability of transition out of the post-holiday state after a stay of one day. In the UK, there is at least one weekend per year -- the Easter weekend -- with public holidays on the neighbouring Friday and Monday. We include the indicator term to allow the transition probability to differ over such weekends, in accordance with the views of industry experts at NGN. An analogous term is included in the logit probability of transition into the post-holiday state, which we model as
\begin{equation}\label{eq:lambda_23}
\logit \{ \lambda_{2,3}(n, p) \} = \nu_{2,3,1} + \nu_{2,3,2} \mathbb{I}(n = 2).
\end{equation}

\subsection{\label{subsec:demand}Modelling the conditional demand for gas}
Analysis involving an earlier version of the model, where observations were assumed to be conditionally independent given the states, revealed substantial autocorrelation between residuals, causing difficulties in identifying the three unknown states. We therefore model the logarithm $\vect{Y}_t$ of the demand for gas, conditional on the state on the current and previous days, as a first-order vector autoregression, or VAR(1), with density $p(\vect{y}_t | \vect{y}_{t-1}, S_{t-1}=s_{t-1}, S_t=s_t, n_t, p_t, r_t, \vect{w}_t)$ for $t=2,\ldots,T$ and, at time $t=1$,
\begin{equation}\label{eq:init_model}
p(\vect{y}_1 | S_0=j, S_1=s_1, n_1, p_1, r_1, \vect{w}_1) = p(\vect{y}_1 | S_1=s_1, n_1, p_1, r_1, \vect{w}_1).
\end{equation}
We constrain our conditional model for gas demand to be conditionally stationary. That is, we can regard the vectors  $\vect{Y}_t$ as being the results of transforming values from a stationary VAR(1) process. The transformation applied on day $t$ depends on the day of the year, day of the week, composite weather variable and state. Given the relatively short time-scales of interest, such a stationarity assumption is reasonable and, in fact, nonstationarity of a non-negligible degree seems implausible in this application.

The conditional model for $(\vect{Y}_{t} | \vect{Y}_{t-1}=\vect{y}_{t-1}, S_{t-1}=s_{t-1}, S_t=s_t)$ is given by
\begin{equation}\label{eq:y}
\vect{Y}_{t} - \vect{\mu}_t = \matr{\Psi} (\vect{y}_{t-1} - \vect{\mu}_{t-1}) + \vect{\epsilon}_t, \quad \vect{\epsilon}_t \sim \mathrm{N}_2 ( \vect{0} , \matr{\Omega}_t^{-1} )
\end{equation}
for $t=2,\ldots,T$, whilst for the initial distribution of $(\vect{Y}_{1} | S_1=s_1)$ in~\eqref{eq:init_model}, we write
\begin{equation}\label{eq:y_init}
\vect{Y}_{1} = \vect{\mu}_{1} + \vect{\epsilon}_1, \quad \vect{\epsilon}_1 \sim \mathrm{N}_2 \left( \vect{0}, V(\matr{\Psi}, \matr{\Omega}_1) \right).
\end{equation}
In both cases the dependence on the states comes through both the time-dependent mean $\vect{\mu}_{t}$ and precision matrix $\matr{\Omega}_t$. The terms $\vect{\epsilon}_1,\ldots,\vect{\epsilon}_T$ form a sequence of independent bivariate normal random vectors with zero mean, $\matr{\Psi} \in \mathcal{S}_{\psi}$ is a $(2 \times 2)$ real-valued matrix, with elements $\Psi_{j,k}$, constrained to satisfy the stationarity condition of a bivariate AR(1) process, and each $\matr{\Omega}_t$ is a $(2 \times 2)$ symmetric, positive definite matrix. The stationarity region $\mathcal{S}_{\psi}$ is defined as the subset of $(2 \times 2)$ matrices with real-valued entries whose eigenvalues are less than 1 in modulus \citep[][]{TWRH16}. The term $\matr{V}(\matr{\Psi}, \matr{\Omega}_1)$ is the stationary variance of the mean-centered errors $\{ \vect{Y}_{t} - \vect{\mu}_t: t=1,2,\ldots \}$ that would prevail if the precision matrix remained equal to $\matr{\Omega}_1$ at all future times, that is, the matrix $\matr{V}$ which solves the equation $\matr{V} = \matr{\Psi} \matr{V} \matr{\Psi}' + \matr{\Omega}_1^{-1}$. A solution, which is symmetric and positive definite, is guaranteed by our assumptions on the support of $\matr{\Psi}$ and $\matr{\Omega}_1$. Specification of a well-behaved prior distribution over the stationarity region $\mathcal{S}_{\psi}$ is very challenging due to its complex geometric constraints. Fortunately, the problem simplifies greatly if we assume $\Psi_{1,1} = \Psi_{2,2} = \Psi_{\text{on}}$ and $\Psi_{1,2} = \Psi_{2,1} = \Psi_{\text{off}}$, in which case the necessary and sufficient conditions for stationarity are that $| \Psi_{\text{on}} + \Psi_{\text{off}} | < 1$ and $| \Psi_{\text{on}} - \Psi_{\text{off}} | < 1$. We therefore adopt this simplification to the model.

The time-dependent mean, $\vect{\mu}_{t}$, in~\eqref{eq:y} and \eqref{eq:y_init} includes terms to allow for the influence of the composite weather variable (CWV) and various seasonal and calendar related effects. As we are working on the logarithmic scale, each has a multiplicative effect on the demand for gas. Conditional on the state, $S_t = s_t$, the mean for LDZ $j$, is given by
\begin{equation}\label{eq:mean}
\mu_{t,j} = \alpha_j + B_{t,j,s_t} \beta_{j,r_t} + \Gamma_{t,j} + \Delta_{t,j} + (\zeta_{j,1} + \zeta_{j,2} w_{t,j}) \tilde{w}_{t,j}
\end{equation}
for $j=1,2$, where $\alpha_j$ provides an intercept. Due to interactions with the temperature and differences in consumer habits, the effect of a public holiday may differ according to whether it falls over the Christmas period, at Easter or over the summer. We therefore allow the three types of public holiday, defined earlier in Table~\ref{tab:bank_hols}, to have  different effects on $\mu_{t,j}$, represented by $\beta_{j,1}$, $\beta_{j,2}$ and $\beta_{j,3}$. The mean depends on the state on day $t$ through the term $B_{t,j,s_t}$ which controls whether or not a holiday effect is included. It is defined by
\begin{equation*}
B_{t,j,1} = \rho_{\beta,j}^{n_t}, \qquad B_{t,j,2} = 1, \qquad B_{t,j,3} = \rho_{\beta,j}^{\text{min}(n_t, p_t)}, \qquad B_{t,j,4} = 0,
\end{equation*}
where $\rho_{\beta,j} \in (0, 1)$ so that the holiday effect $\beta_{j,r_t}$ is scaled down on proximity days (states 1 and 3) by a factor which decays to zero with increasing separation. Since the other terms in~\eqref{eq:mean} do not depend on the state on day $t$, the mean $\mu_{t,j}$  in the pre- or post-holiday state is a weighted average of the mean in the normal and holiday states. The weights depend on the number of days separating day $t$ from the next or closest public holiday, respectively.

In~\eqref{eq:mean}, the covariate $\tilde{w}_{t,j}$ is defined by $\tilde{w}_{t,j} = w_{t,j} - m_{d(t),j}$, where $m_{d(t),j}$ is a smoothed average for the CWV in region $j$ on day $t$ where $d(t) \in \{ 1, 2, \ldots, 366\}$. We include both the mean-centered CWV, $\tilde{w}_{t,j}$, and its interaction with the raw CWV, $w_{t,j}$, to allow the effect of above or below average temperatures to differ according to absolute weather conditions. The structure of this term is supported by Supplementary Figure~\ref{fig:cencwv_vs_logdemand} which shows the effect of $w_{t,j}$ on the relationship between log gas demand and the mean-centered CWV. 

The term $\Delta_{t,j}$ is constructed to give a day of the week effect, whilst $\Gamma_{t,j}$ provides a seasonal component, after allowing for the CWV. Both terms are composed using Fourier series with
\begin{equation}\label{eq:zeta}
\Delta_{t,j} = \sum_{k=1}^3 \left\{ \delta_{j,1,k} \cos\left( \frac{2 \pi k t}{7} \right) + \delta_{j,2,k} \sin\left( \frac{2 \pi k t}{7} \right) \right\}
\end{equation}
and
\begin{equation}\label{eq:xi}
\Gamma_{t,j} = \sum_{k=1}^{\mathrm{K}_{\gamma}} \left\{ \gamma_{j,1,k} \cos\left( \frac{2 \pi k t}{365.25} \right) + \gamma_{j,2,k} \sin\left( \frac{2 \pi k t}{365.25} \right)  \right\}.
\end{equation}
The six unconstrained parameters $\vect{\delta}_j=(\delta_{j,1,1},\ldots,\delta_{j,2,3})' \in \mathbb{R}^6$ in~\eqref{eq:zeta} provide fixed effects for each day of the week which, by construction, sum to zero. The advantages of this parameterisation are that the six elements of $\vect{\delta}_j$ are unconstrained and by treating them as exchangeable \emph{a priori}, we can induce a prior for the seven fixed effects which is symmetric with respect to the day of the week. The Fourier series for the seasonal term in~\eqref{eq:xi} is truncated at a modest number, $\mathrm{K}_{\gamma}$, of harmonics beyond which we assume their contribution to be negligible. The choice of $\mathrm{K}_{\gamma}$ will be discussed further in Section~\ref{sec:appl}.

The time-dependent precision matrix $\matr{\Omega}_t$ in~\eqref{eq:y} and \eqref{eq:y_init} also depends on $S_t$, with different values for the holiday and normal states when $S_t=2$ and $S_t=4$, respectively. Adopting an approach similar to that taken for the mean $\vect{\mu}_t$, we model the precision matrix on proximity days (states 1 and 3) by interpolating between these two values, with weights that depend on the number of days to or from the neighbouring public holiday. This is most natural when working in $\mathbb{R}^3$, rather than the constrained space of $2 \times 2$ covariance matrices. We therefore reparameterise the precision matrix $\matr{\Omega}_t$ in terms of its square--root--free Cholesky decomposition $\matr{\Omega}_t = \matr{T}_{\Omega_t}' \matr{D}_{\Omega_t}^{-1} \matr{T}_{\Omega_t}$ \citep[][]{Pou99}, in which $\matr{T}_{\Omega_t}$ is a unit lower triangular matrix with $(2,1)$-element $- \phi_t$ and $\matr{D}_{\Omega_t} = \text{diag}(\tau_{t,1}^{-1}, \tau_{t,2}^{-1})$. The new parameters have a convenient interpretation in terms of the autoregression of $\epsilon_{t,2}$ on $\epsilon_{t,1}$, with $\phi_t \in \mathbb{R}$ representing the autoregressive coefficient and $\tau_{t,2} > 0$ the associated conditional precision. The parameter $\tau_{t,1} > 0$ represents the marginal precision of $\epsilon_{t,1}$. For the coefficient, $\omega_{t,1} = \phi_t$, and the log precisions, $\omega_{t,2} = \ln (\tau_{t,1})$ and $\omega_{t,3} = \ln (\tau_{t,2})$, we adopt linear models with, for $i=1,2,3$,
\begin{equation}\label{eq:prec}
\omega_{t,i} = \eta_i + \Theta_{t,i,s_t} \theta_i + K_{t,i},
\end{equation}
 in which $\eta_i$ is an intercept and $\theta_i$ allows for the effect of a public holiday. Again, we assume that days before and after a holiday have the same effect and define
\begin{equation*}
\Theta_{t,i,1} = \rho_{\theta}^{n_t}, \qquad \Theta_{t,i,2} = 1, \qquad \Theta_{t,i,3} = \rho_{\theta}^{\text{min}(n_t, p_t)}, \qquad \Theta_{t,i,4} = 0,
\end{equation*}
where $\rho_{\theta} \in (0, 1)$, which scales down the holiday effect $\theta_i$ on proximity days. The term $K_{t,i}$ in~\eqref{eq:prec} provides a seasonal component. The latter was deemed necessary after analysis with an earlier version of the model, in which $K_{t,i}$ was omitted, revealed seasonal variation in the residual variance, which caused confounding between the state allocation and low frequency seasonal change. As with the seasonal component $\Gamma_{t,j}$ in the time-varying mean, we use a truncated Fourier series to represent $K_{t,i}$, taking
\begin{equation}\label{eq:H}
K_{t,i} = \sum_{k=1}^{\mathrm{K}_{\kappa}} \left\{ \kappa_{i,1,k} \cos\left( \frac{2 \pi k t}{365.25} \right) + \kappa_{i,2,k} \sin\left( \frac{2 \pi k t}{365.25} \right)  \right\}.
\end{equation}
The choice of truncation point, $\mathrm{K}_{\kappa}$, will be discussed further in Section~\ref{sec:appl}.

\section{\label{sec:prior}Prior distribution}
Denoting the unknown parameters of the Markov model for state evolution by $\matr{\Lambda}$ and the parameters of the conditional model for demand by $\matr{\Pi}$, we adopt a prior distribution in which  $\matr{\Lambda}$ and $\matr{\Pi}$ are independent. The two independent components of our prior are described in the sections which follow. 

\subsection{\label{subsec:prior_lambda}Transition probabilities}
The parameters $\matr{\Lambda} = (\nu_{4,1,1},\nu_{4,1,2},\nu_{3,4,1},\nu_{3,4,2},\nu_{3,4,3},\nu_{2,3,1},\nu_{2,3,2})'$ comprise the linear coefficients from the logit probabilities~\eqref{eq:lambda_41}--\eqref{eq:lambda_23}. We adopt a prior with independence between the $\nu_{j,k,i}$ and take
\begin{equation}\label{eq:prior_lambda}
\nu_{j,k,i} \sim \mathrm{N}(m_{j,k,i}, v_{j,k,i}).
\end{equation}
The symmetric logit normal distribution, created by assigning a normal $\mathrm{N}(0, v)$ distribution to the logit transformation of a probability-valued random variable, becomes bimodal when $v > 1$. U-shaped priors for the transition probabilities $\lambda_{j,k}(n, p)$ were not consistent with our prior beliefs and so we moderated our choice of prior variances $v_{j,k,i}$ to avoid bimodality. In order to avoid the complicated likelihood leading to implausible sets of parameter values having undue weight in the posterior, a process of trial and improvement, guided by the interpretation of the $\nu_{j,k,i}$, was then used to choose the hyperparameters. For a recent discussion of such issues see, for example, \citet{GSB17}.

\subsection{\label{subsec:prior_theta}Parameters of the conditional demand model}
\begin{sloppypar}
For region $j=1,2$, let $\vect{\beta}_{j}=(\beta_{j,1},\ldots,\beta_{j,3})'$, $\vect{\zeta}_{j}=(\zeta_{j,1},\zeta_{j,2})'$, $\vect{\gamma}_j = (\gamma_{j,1,1}, \ldots, \gamma_{j,1,\mathrm{K}_{\gamma}}, \gamma_{j,2,1}, \ldots, \gamma_{j,2,\mathrm{K}_{\gamma}})'$ and, similarly, $\vect{\delta}_{j} = (\delta_{j,1,1}, \ldots, \delta_{j,2,3})'$. Define $\vect{\beta} = (\vect{\beta}_1', \vect{\beta}_2')'$, with corresponding concatenated vectors for the other region-specific parameters. In the model for the time-varying precision matrix, let $\vect{\eta}=(\eta_1, \eta_2, \eta_3)'$, $\vect{\theta}=(\theta_1, \theta_2, \theta_3)'$, $\vect{\kappa}_i=(\kappa_{i,1,1}, \ldots, \kappa_{i,2,\mathrm{K}_{\kappa}})'$ and $\vect{\kappa} = (\vect{\kappa}_1', \vect{\kappa}_2', \vect{\kappa}_3')'$. Finally, let $\vect{\rho}=(\rho_{\beta,1}, \rho_{\beta,2}, \rho_{\theta})'$ for the parameters governing the rate of decay of the holiday effects in the time-varying mean and precision matrix. The model parameters then comprise $\matr{\Pi} = \{ \vect{\Psi}, \vect{\alpha}, \vect{\beta}, \vect{\gamma}, \vect{\delta}, \vect{\rho}, \vect{\zeta}, \vect{\eta}, \vect{\theta}, \vect{\kappa} \}$. We impose prior independence between these parameter blocks. 
\end{sloppypar}

As discussed in Section~\ref{subsec:demand} we assume the autoregressive coefficient matrix $\matr{\Psi}$ is composed of a common diagonal element $\Psi_{\text{on}}$ and a common off-diagonal element $\Psi_{\text{off}}$ so that the stationarity condition reduces to $| \Psi_{\text{on}} + \Psi_{\text{off}} | < 1$ and $| \Psi_{\text{on}} - \Psi_{\text{off}} | < 1$. To make it easier to impose this constraint, we first define $\chi_1 = \Psi_{\text{on}} + \Psi_{\text{off}}$ and $\chi_2 = \Psi_{\text{on}} - \Psi_{\text{off}}$ and then reparameterise $\matr{\Psi}$ in terms of new parameters $\xi_i = (\chi_i + 1) / 2 \in (0, 1)$ for $i=1,2$. To construct our prior, we make $\xi_1$ and $\xi_2$  independent and give them beta distributions, $\xi_i \sim \mathrm{Beta}(a_{\xi,i}, b_{\xi,i})$.

For the parameters in the time-varying mean $\vect{\mu}_t$ governing the intercept, influence of the CWV, and the seasonal, day of the week and public holiday effects, we adopt hierarchical priors which allow information to be shared between the NO ($j=1$) and NE ($j=2$) LDZs. For the intercept we take
$\alpha_{j} \mid \mu_{\alpha} \stackrel{\text{\scriptsize{\textit{i.i.d}}}}{\sim} \mathrm{N}(\mu_{\alpha}, [1-r_{\alpha}] v_{\alpha} )$, for $ j=1,2$, with $ \mu_{\alpha} \sim \mathrm{N}(m_{\alpha}, r_{\gamma} v_{\alpha})$, 
in which the correlation between sites is $r_{\alpha}$, the marginal variance is $v_{\alpha}$ and the marginal mean is $m_{\alpha}$. Similarly, independently for the coefficients of the mean-centred CWV ($i=1$) and its interaction with the raw value ($i=2$), we choose
$\zeta_{j,i} \mid \mu_{\alpha} \stackrel{\text{\scriptsize{\textit{i.i.d}}}}{\sim} \mathrm{N}(\mu_{\zeta,i}, [1-r_{\zeta,i}] v_{\zeta,i} )$, for $ j=1,2$, with $ \mu_{\zeta,i} \sim \mathrm{N}(m_{\zeta,i}, r_{\zeta,i} v_{\zeta,i})$.
For the day of the week effects, independently for $m=1,2$ and $k=1,2,3$ we choose
$\delta_{j,m,k} \mid \mu_{\delta,m,k} \stackrel{\text{\scriptsize{\textit{i.i.d}}}}{\sim} \mathrm{N}(\mu_{\delta,m,k}, [1-r_{\delta}] v_{\delta} )$, for $j=1,2$, with $ \mu_{\delta,m,k} \sim \mathrm{N}(0, r_{\delta} v_{\delta})$.
For the seasonal effects, independently for $m=1,2$ and $k=1,\ldots,\mathrm{K}_{\gamma}$, we take
$\gamma_{j,m,k} \mid \mu_{\gamma,m,k} \stackrel{\text{\scriptsize{\textit{i.i.d}}}}{\sim} \mathrm{N}(\mu_{\gamma,m,k}, [1-r_{\gamma}] v_{\gamma,k} )$, for $ j=1,2$, with $ \mu_{\gamma,m,k} \sim \mathrm{N}(0, r_{\gamma} v_{\gamma,k})$,
in which $v_{\gamma,1} \ge v_{\gamma,2} \ge \ldots \ge v_{\gamma,\mathrm{K}_{\gamma}}$. We note that the assignment of independent, zero mean normal distributions, $\mathrm{N}(0, v_{\gamma,k})$, to a pair of Fourier coefficients, $\gamma_{j,1,k}$ and $\gamma_{j,2,k}$, is equivalent to assigning a uniform distribution to the phase of the $k$-th harmonic and, independently, a Rayleigh distribution to its amplitude, with scale parameter $\sqrt{v_{\gamma,k}}$. Our prior therefore conveys the idea that the size of the seasonal harmonics will decay as their frequency increases.

For the public holiday effects, we adopt a prior of the form
$\vect{\beta}_{j} \mid \vect{\mu}_{\beta} \stackrel{\text{\scriptsize{\textit{i.i.d}}}}{\sim} \mathrm{N}_3 ( \vect{\mu}_{\beta}, [1-r_{\beta}] \vect{V}_{\beta} )$, For $ j=1,2$, with $ \vect{\mu}_{\beta} \sim \mathrm{N}_3( \vect{0}, r_{\beta} \vect{V}_{\beta})$.
Here $\vect{V}_{\beta}$ is a $3 \times 3$ compound symmetric matrix with non-zero off-diagonal elements, allowing positive correlation between the effects of each type of public holiday so that information can be pooled across types. This offers a compromise between an impractical assumption that the $\beta_{j,i}$ are independent for each site $j$, which would limit the information available for inference, and an inflexible assumption that they are all equal. 

The factor $\rho_{\beta,j}$ acts on proximity days to scale down the holiday effect in the time-varying mean $\mu_{t,j}$ for LDZ $j$, $j=1,2$. Similarly, the factor $\rho_{\theta}$ scales down the holiday effect in our reparameterisation of the time-varying precision matrix, $\matr{\Omega}_t$, of the errors $\vect{\epsilon}_t$. The factors $\rho_{\beta,1}$ and $\rho_{\beta,2}$ in the means for the Northern and North East LDZs have the same function and we expect them to be very similar. Although we would also expect the $\rho_{\beta,j}$ to be informative about $\rho_{\theta}$, we believe that they will carry less information because they relate to the rate of decay of parameters which play different roles. We therefore construct an asymmetric hierarchical prior by taking $\tilde{\rho}_{\cdot} = \logit(\rho_{\cdot}) = \ln\{\rho_{\cdot}/(1-\rho_{\cdot})\}$, and then choosing
$\tilde{\rho}_{\beta,j} \mid \tilde{\rho}_{\beta} \stackrel{\text{\scriptsize{\textit{i.i.d}}}}{\sim} \mathrm{N}(\tilde{\rho}_{\beta}, [1-r_{\tilde{\rho},1}] v_{\tilde{\rho}} ) $
in the means for LDZs $j=1,2$ and then
$\tilde{\rho}_{\beta} \mid \mu_{\tilde{\rho}} \sim \mathrm{N}(\mu_{\tilde{\rho}}, [1-r_{\tilde{\rho},2}] r_{\tilde{\rho},1} v_{\tilde{\rho}} )$  and $ \tilde{\rho}_{\theta} \mid \mu_{\tilde{\rho}} \sim \mathrm{N}(\mu_{\tilde{\rho}}, [1-r_{\tilde{\rho},2}] r_{\tilde{\rho},1} v_{\tilde{\rho}} )$ 
and finally
$\mu_{\tilde{\rho}} \sim \mathrm{N}(m_{\tilde{\rho}}, r_{\tilde{\rho},1} r_{\tilde{\rho},2} v_{\tilde{\rho}})$.
The marginal prior moments for the logit parameters are $\E(\tilde{\rho}_{\beta,j}) = \E(\tilde{\rho}_{\theta}) = m_{\tilde{\rho}}$, $\Var(\tilde{\rho}_{\beta,j}) = v_{\tilde{\rho}}$, $\Cor(\tilde{\rho}_{\beta,j}, \tilde{\rho}_{\theta}) = \sqrt{r_{\tilde{\rho},1}} r_{\tilde{\rho},2}$, for $j=1,2$, and $\Var(\tilde{\rho}_{\theta}) = r_{\tilde{\rho},1} v_{\tilde{\rho}}$, $\Cor(\tilde{\rho}_{\beta,1}, \tilde{\rho}_{\beta,2}) = r_{\tilde{\rho},1}$ where $r_{\tilde{\rho},1}, r_{\tilde{\rho},2} \in (0, 1)$. We can therefore choose the correlation between the $\tilde{\rho}_{\beta,j}$ to be greater than their correlation with $\tilde{\rho}_{\theta}$. 

Finally, the intercept and holiday effects in our reparameterised time-varying precision matrix are given independent normal distributions, taking $\eta_{i} \sim \mathrm{N}( m_{\eta,i}, v_{\eta,i} )$ and $\theta_i \sim \mathrm{N}(0, v_{\theta,i})$ for $i=1,2,3$. Similarly, the seasonal effects are given independent normal priors, with $\kappa_{i,m,k} \sim \mathrm{N}(0, v_{\kappa,i,k})$, $m=1,2$, $k=1,\ldots,\mathrm{K}_{\kappa}$, for $i=1,2,3$.

Our choices for the hyperparameters in each component of our prior distribution are detailed in Section~\ref{subsec:prior_spec}. 

\section{\label{sec:post_inf}Posterior inference}
The posterior distribution of the model parameters follows from Bayes Theorem as
\begin{equation*}
\pi(\matr{\Pi}, \matr{\Lambda} | \vect{y}, \vect{n}, \vect{p}, \vect{r}, \vect{w}) \propto p(\vect{y} | \matr{\Pi}, \matr{\Lambda}, \vect{n}, \vect{p}, \vect{r}, \vect{w}) \pi(\matr{\Pi}) \pi(\matr{\Lambda}) 
\end{equation*}
in which $\vect{y}$ represents the complete time series, here $\vect{y} = \vect{y}_{1:T}$, where the general notation $x_{i:j}$ for $i < j$ indicates a sequence $(x_i, x_{i+1}, \ldots, x_{j-1}, x_j)$. The term $p(\vect{y} | \matr{\Pi}, \matr{\Lambda}, \vect{n}, \vect{p}, \vect{r}, \vect{w})$ is the observed data likelihood given by
\begin{equation*}
p(\vect{y} | \matr{\Pi}, \matr{\Lambda}, \vect{n}, \vect{p}, \vect{r}, \vect{w}) = \sum_{\vect{s}} p(\vect{y} | \vect{s}, \matr{\Pi}, \vect{n}, \vect{p}, \vect{r}, \vect{w}) p(\vect{s} | \matr{\Lambda}, \vect{n}, \vect{p})
\end{equation*}
in which the sum is taken over all possible sequences of states, $\vect{s}=\vect{s}_{0:T}$. It can be calculated efficiently using a filtering algorithm which computes $\Pr(S_t = k, \vect{y}_{1:t} | \matr{\Pi}, \matr{\Lambda}, \vect{n}, \vect{p}, \vect{r}, \vect{w})$, $k \in \mathcal{S}_s$, recursively for $t=0,\ldots,T$ and finally
\begin{equation*}
p(\vect{y} | \matr{\Pi}, \matr{\Lambda}, \vect{n}, \vect{p}, \vect{r}, \vect{w}) = \sum_{k=1}^4 \Pr(S_T = k, \vect{y}_{1:T} | \matr{\Pi}, \matr{\Lambda}, \vect{n}, \vect{p}, \vect{r}, \vect{w}).
\end{equation*}
The lagged dependence of $\vect{Y}_{t+1}$ on $S_t$ in the DAG in Figure~\ref{fig:DAG} complicates recursive algorithms for quantifying posterior uncertainty in the hidden states. However, the model can be reformulated to simplify its conditional independence structure by defining an augmented state on day $t$ by $\tilde{S}_t = (S_{t-1}, S_t)'$ for $t=1,\ldots,T$. The demand  $\vect{Y}_t$ is then conditionally independent of the augmented state on the previous day $\tilde{S}_{t-1}$ given the augmented state on the current day $\tilde{S}_t$ and the previous observation $\vect{Y}_{t-1}$. As some transitions in our original state process have zero probability, there are only $11$, rather than $4^2=16$, possible values for the augmented state $\tilde{S}_t$ and we denote this set of values by $\mathcal{S}_{\tilde{s}}$. The mapping from $(\mathcal{S}_s \times \mathcal{S}_s)$ to $\mathcal{S}_{\tilde{s}}$ is detailed in Table~\ref{tab:def_aug} of the Supplementary Materials. The transition matrix for the new hidden process is sparse and can be represented in terms of the original transition probabilities $\lambda_{j,k}(n, p)$, as indicated in Supplementary Table~\ref{tab:trans}. The initial distribution $\Pr(\tilde{S}_1 | \matr{\Lambda}, \vect{p}_{0:1}, \vect{n}_{0:1})$ can similarly be formed by taking an appropriate product of terms, $\ell_j(n_0, p_0)$ and $\lambda_{j,k}(n_1, p_1)$. Filtering algorithms for HMMs of this simple form are standard. See, for example, Chapter 2 of \citet{MZ97}. Our algorithm is given in Section~\ref{subsec:forward} of the Supplementary Materials.

The posterior distribution $\pi(\matr{\Pi}, \matr{\Lambda} | \vect{y}, \vect{n}, \vect{p}, \vect{r}, \vect{w})$ cannot be evaluated in closed form and so we build a numerical approximation using a Hamiltonian Monte Carlo sampling scheme. We note that the likelihood $p(\vect{y} | \matr{\Pi}, \matr{\Lambda}, \vect{n}, \vect{p}, \vect{r}, \vect{w})$ is not symmetric with respect to the labels of the hidden states. As a result the state-specific parameters are all identifiable in the posterior, and so methods to force an identified sample, such as those discussed in \citet{Ste00}, are not required.

\subsection{Approximating the posterior for the model parameters}
Hamiltonian Monte Carlo (HMC) is a special case of the Metropolis algorithm; for example, see \citet{Nea11} or \citet{Bet17} for an introduction. It relies on the introduction of auxiliary variables that are interpreted as the momentum of a particle whose position in space is represented by the parameter values. This allows efficient proposals to be generated by exploiting Hamiltonian dynamics and modelling the movement of the sampler around the joint posterior of the momentum and position variables as the motion of the particle through an unbounded, frictionless space. We use \texttt{rstan} \citep[][]{Sta16b}, the R interface to the Stan software \citep[][]{CGH17}, to implement the HMC algorithm. Stan requires users to write a program in the probabilistic Stan modelling language, the role of which is to provide instructions for computing the logarithm of the kernel of the posterior density function. The Stan software then automatically tunes and runs a Markov chain simulation to sample from the resulting posterior.

\subsection{Approximating the posterior for the hidden states}
Although the hidden states $S_t$ are not sampled as part of the HMC scheme, their marginal posteriors can be approximated by Rao-Blackwellisation through:
\begin{equation}\label{eq:smoothed_probs}
\hat{\Pr}(S_t = k | \vect{y}, \vect{n}, \vect{p}, \vect{r}, \vect{w}) = \frac{1}{M} \sum_{i=1}^M \Pr(S_t = k | \vect{y}, \matr{\Pi}^{[i]}, \matr{\Lambda}^{[i]}, \vect{n}, \vect{p}, \vect{r}, \vect{w}), \quad k \in \mathcal{S}_{s},
\end{equation}
for $t=0,\ldots,T$ where $\matr{\Pi}^{[i]}$ and $\matr{\Lambda}^{[i]}$ denote the $i$-th posterior samples of the model parameters $\matr{\Pi}$ and $\matr{\Lambda}$ for $i=1,\ldots,M$. The full sample smoothed probabilities $\Pr(\tilde{S}_t = k | \vect{y}, \matr{\Pi}, \matr{\Lambda}, \vect{n}, \vect{p}, \vect{r}, \vect{w})$, can be computed in a backward recursion, starting at time $t=T$, using the forward probabilities $\Pr(\tilde{S}_t = k, \vect{y}_{1:t} | \matr{\Pi}, \matr{\Lambda}, \vect{n}, \vect{p}, \vect{r}, \vect{w})$ calculated when computing the observed data likelihood. We can then marginalise over $S_{t-1}$ to compute the probabilities in our original four-state model $\Pr(S_t = k | \vect{y}, \matr{\Pi}, \matr{\Lambda}, \vect{n}, \vect{p}, \vect{r}, \vect{w})$, $k \in \mathcal{S}_s$ for use in~\eqref{eq:smoothed_probs}. The complete algorithm is given in Section~\ref{subsec:backward} of the Supplementary Materials.

\section{\label{sec:appl}Application to daily demand data}
The daily gas consumption data from NGN were introduced in Section~\ref{sec:data}. Using the algorithm described in Section~\ref{sec:post_inf}, we fitted our hierarchical model to data from each of the three NDM load bands. When including a large number of Fourier components in the model for mean log demand, the amplitude, $(\gamma_{j,1,k}^2 + \gamma_{j,2,k}^2)^{1/2}$, of the $k$-th harmonic in~\eqref{eq:xi} for each site $j$ was negligible when $k>6$ and so we truncated the series at $\mathrm{K}_{\gamma}=6$. Applying similar reasoning, we truncated the Fourier series in~\eqref{eq:H} for the seasonally varying precision at $\mathrm{K}_{\kappa}=12$. 

\subsection{\label{subsec:prior_spec}Prior specification}
The structure of the prior distribution for the model parameters was outlined in Section~\ref{sec:prior}. It began with an assumption of independence between the parameters $\matr{\Lambda}$ of the marginal model for the hidden states and the parameters $\matr{\Pi}$ of the conditional model for gas demand given the states. The moments in the prior for $\matr{\Pi}$ were chosen according to the specification in Section~\ref{subsec:prior_demand} of the Supplementary Materials and the densities were generally very flat.

Completing our prior for $\matr{\Lambda}$ requires choices of  means and variances in the normal distributions~\eqref{eq:prior_lambda} for the logit coefficients. Our general procedure was described in Section~\ref{subsec:prior_lambda}. By several iterations of calculation and inspection we arrived at the specification in Section~\ref{subsec:prior_tp} of the Supplementary Materials which yielded the distribution for the states shown in Figure~\ref{fig:s_post_2015_LB3} over a representative year. The pointwise prior mode for the entire state sequence is displayed in Supplementary Figure~\ref{fig:s_prior_mode}.

\subsection{\label{subsec:hmc_imp}HMC implementation}
The Stan program representing our hierarchical model is available in the Supplementary Materials. For the data from each load band, using the \texttt{rstan} interface to the Stan software, we ran four HMC chains initialised at different starting points for 10000 iterations, half of which were discarded as burn-in. It was convenient to compute the posterior samples for the smoothed probabilities $\Pr(S_t = \ell | \vect{y}, \matr{\Pi}, \matr{\Lambda}, \vect{n}, \vect{p}, \vect{r}, \vect{w})$ in~\eqref{eq:smoothed_probs} online. To reduce the associated storage overheads, we therefore thinned the output to retain every 10th posterior draw. The usual graphical and numerical diagnostics gave no evidence of any lack of convergence and the effective sample size was at least 1565 for every parameter.

\subsection{\label{subsec:post_inf}Posterior inference}

\subsubsection{Parameter inference}

\begin{figure}[t]
\centering
\scalebox{0.7}{\includegraphics{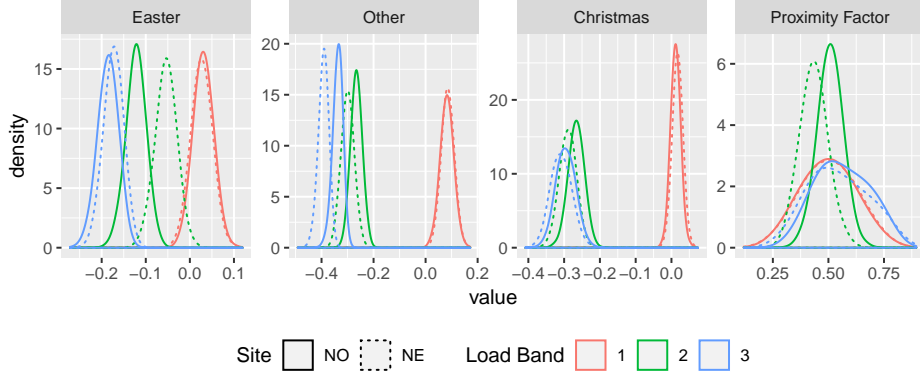}}
\caption{\label{fig:hol_effects_mean}Panels 1--3: marginal posterior densities for the fixed effects $\beta_{j,k}$ in the mean $\mu_{t,j}$ for each type, $k=1,2,3$, of public holiday in each LDZ, $j=1,2$. Panel 4: marginal posterior density for the parameter $\rho_{\beta,j}$ governing the rate of decay of the holiday effect on proximity days in each LDZ, $j=1,2$. Densities for all load bands are shown.}
\end{figure}

Figure~\ref{fig:hol_effects_mean} shows the fixed effects $\beta_{j,k}$ in the mean $\mu_{t,j}$ \eqref{eq:mean} for each type of public holiday in each LDZ. As expected, there is evidence that the mean takes larger values on public holidays in load band 1, which comprises domestic customers, and smaller values in load bands 2 and 3, which represent commercial and small industrial customers, whose business activity typically stops on holidays. The effect seems to be the largest in magnitude amongst industrial customers (load band 3) and smallest amongst domestic users. In general, the effect of a public holiday is less pronounced during the public holidays around Easter. The right-hand panel of Figure~\ref{fig:hol_effects_mean} shows the marginal posterior density for the parameter $\rho_{\beta,j}$ which governs the rate of decay of the holiday effect in the pre- and post-holiday states. The posteriors for each LDZ in load band 3 seems to support larger values than those for load bands 1 and 2, suggesting a slower rate of decay.

\begin{figure}[t]
\centering
\scalebox{0.7}{\includegraphics{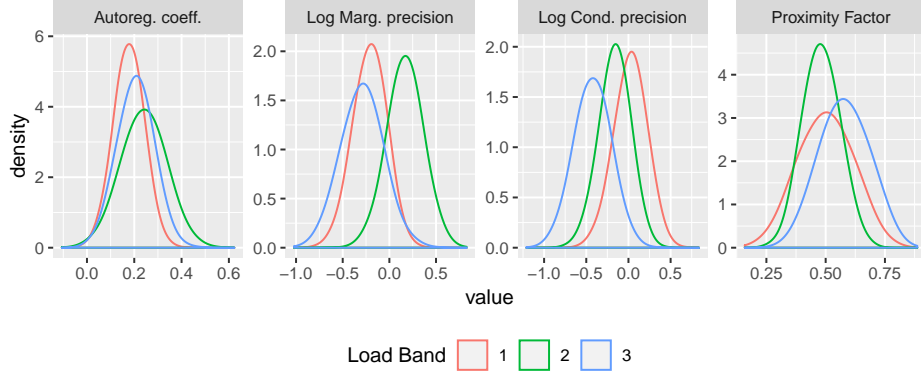}}
\caption{\label{fig:hol_effects_prec}Panels 1--3: marginal posterior densities for the fixed effects $\theta_i$ in the parameters $\omega_{t,i}$ of the square--root free decomposition of the precision matrix $\matr{\Omega}_t$, where $\omega_{t,1} = \phi_t$, $\omega_{t,2} = \ln (\tau_{t,1})$ and $\omega_{t,3} = \ln (\tau_{t,2})$. Panel 4: marginal posterior density for the parameter $\rho_{\theta}$ governing the rate of decay of the holiday effect on proximity days. Densities for all load bands are shown.}
\end{figure}

The corresponding plots for the fixed effects $\theta_i$ in the parameters $\omega_{t,i}$ \eqref{eq:prec} of the square-root free Cholesky decomposition of the precision matrix $\matr{\Omega}_t$ are shown in Figure~\ref{fig:hol_effects_prec}. For all three load bands, there is strong evidence of a positive effect on the autoregressive coefficient $\omega_{t,1}=\phi_t$, suggesting a stronger positive correlation between the residuals in the two LDZs on public holidays. For the logarithms of the marginal error precision $\omega_{t,2} = \ln (\tau_{t,1})$ and the conditional error precision $\omega_{t,3} = \ln (\tau_{t,2})$, there is little evidence of a public holiday effect, except for a negative relationship with the log conditional precision in load band 3. This suggests public holidays are associated with higher residual variance for (log) gas demand by industrial customers.

\subsubsection{State inference}
\begin{figure}[t]
\centering
\scalebox{0.7}{\includegraphics{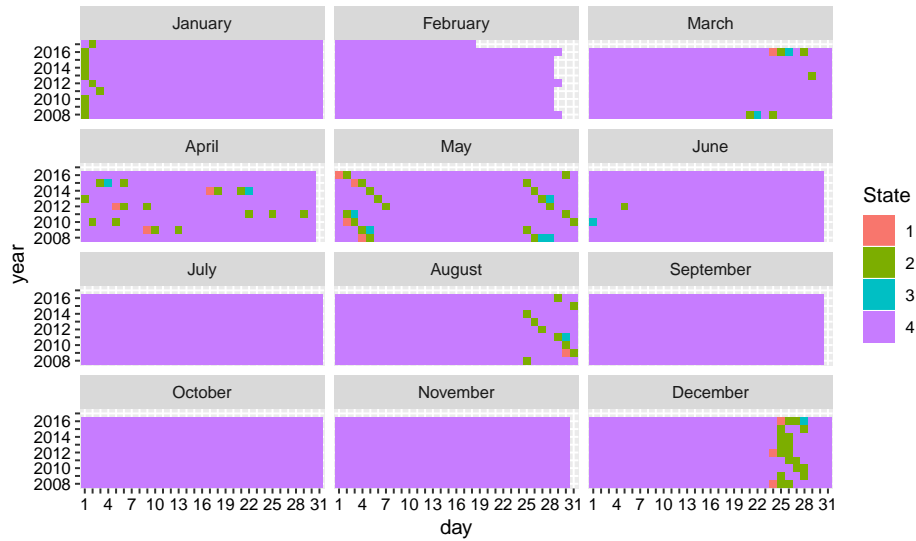}}
\caption{\label{fig:s_post_mode_LB1}Pointwise posterior mode for the state sequence $S_t$ for load band 1.}
\end{figure}

\begin{figure}[t]
\centering
\scalebox{0.7}{\includegraphics{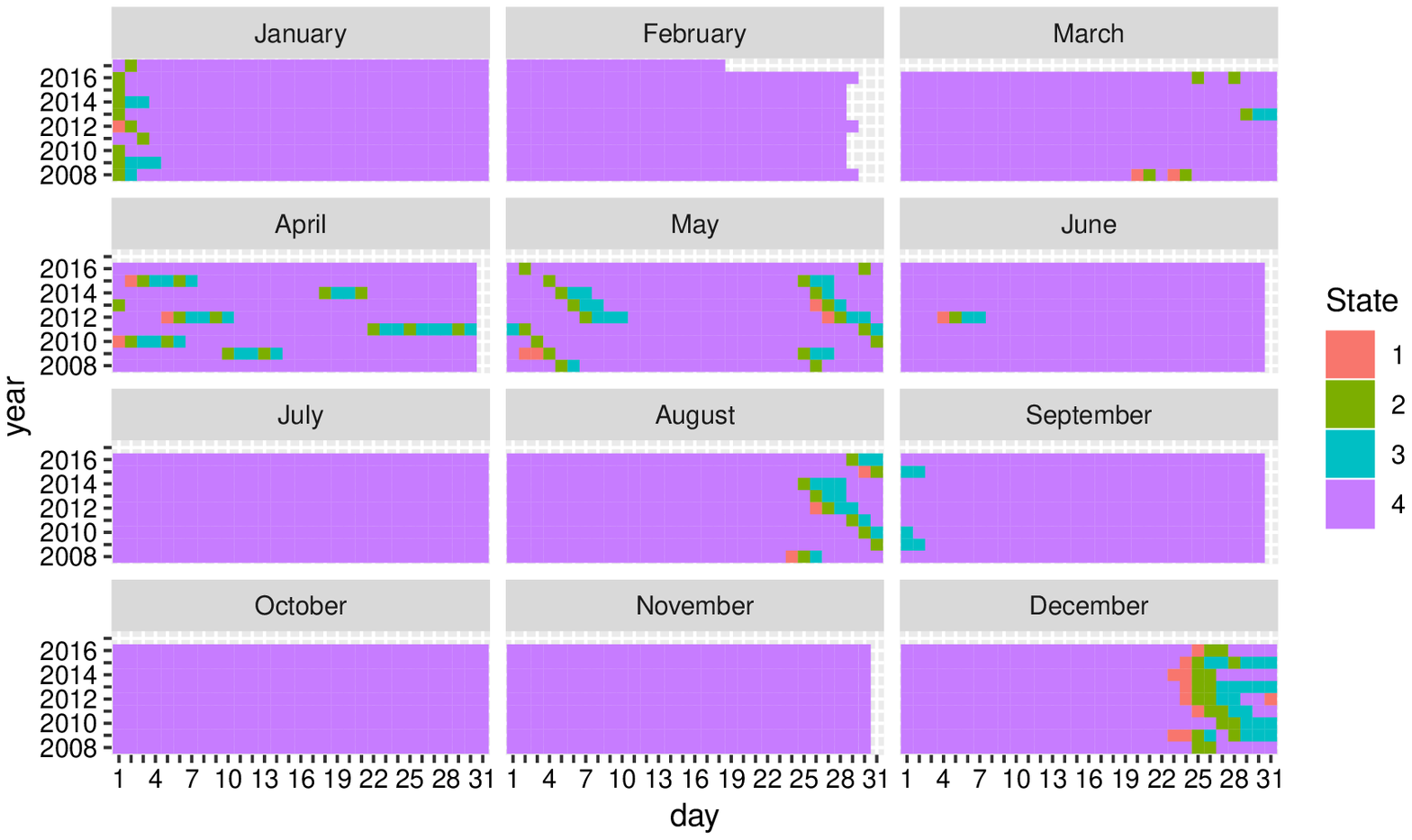}}
\caption{\label{fig:s_post_mode_LB2}Pointwise posterior mode for the state sequence $S_t$ for load band 2.}
\end{figure}

\begin{figure}[t]
\centering
\scalebox{0.7}{\includegraphics{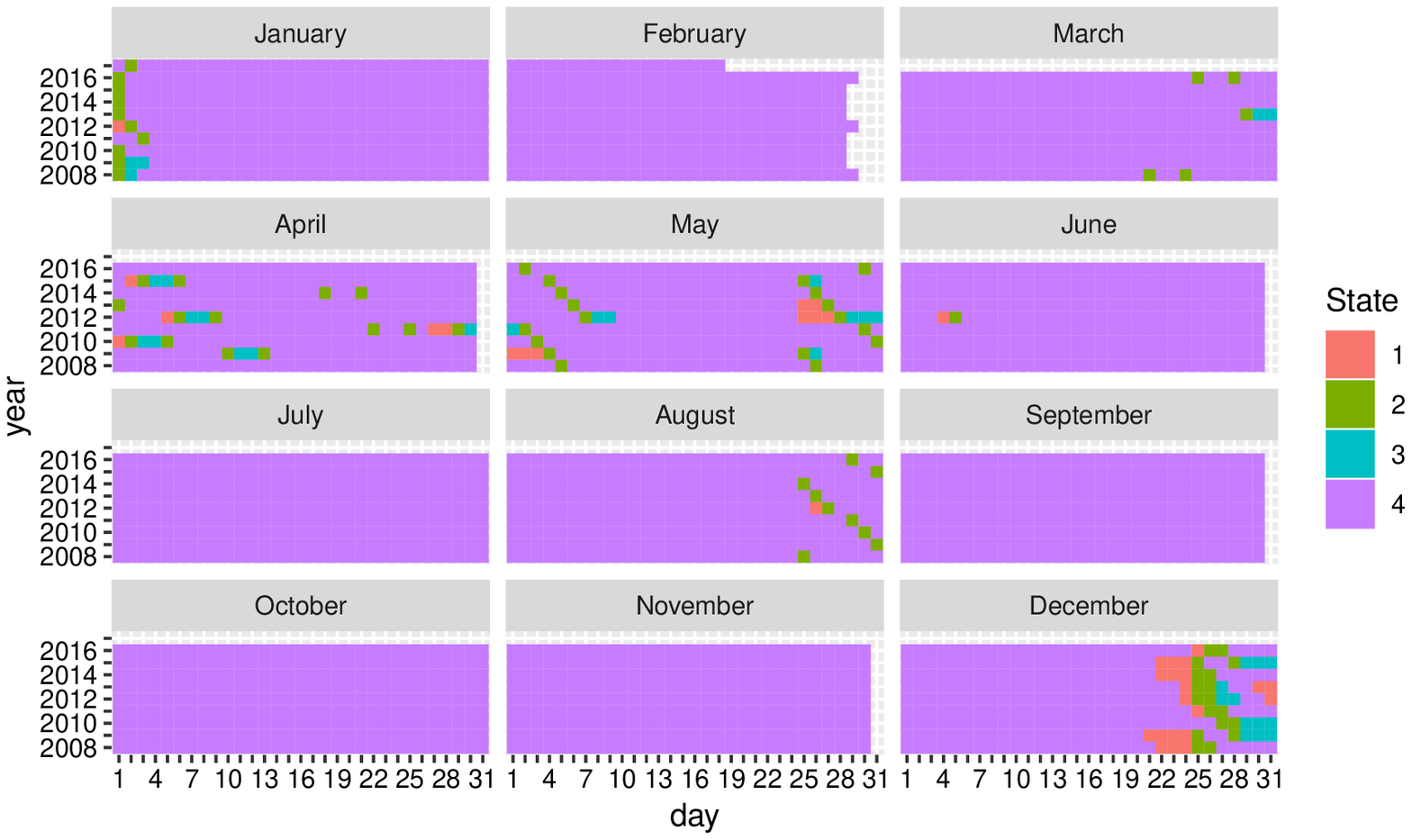}}
\caption{\label{fig:s_post_mode_LB3}Pointwise posterior mode for the state sequence $S_t$ for load band 3.}
\end{figure}

For load bands 1--3, respectively, Figures~\ref{fig:s_post_mode_LB1}--\ref{fig:s_post_mode_LB3} show the pointwise posterior mode for the state sequence, $S_1,\ldots,S_T$. Different patterns are evident across the different load bands. In load band 1, there is little evidence of a proximity effect, with the posterior probability of assignment to the pre- and post-holiday states, 1 and 3, being less than 0.5 in the neighbourhood of most public holidays. For load bands 2 and 3, Figures~\ref{fig:s_post_mode_LB2} and \ref{fig:s_post_mode_LB3} provide clear evidence of a proximity effect around the Christmas period and over the Easter weekend, with additional evidence of a post-holiday effect after the late spring and summer public holidays in load band 2. The sojourn times in the proximity states are longest around Christmas, most notably before Christmas Day in load band 3 and  after Boxing Day in load band 2. 

\begin{figure}[t]
\centering
\scalebox{0.7}{\includegraphics{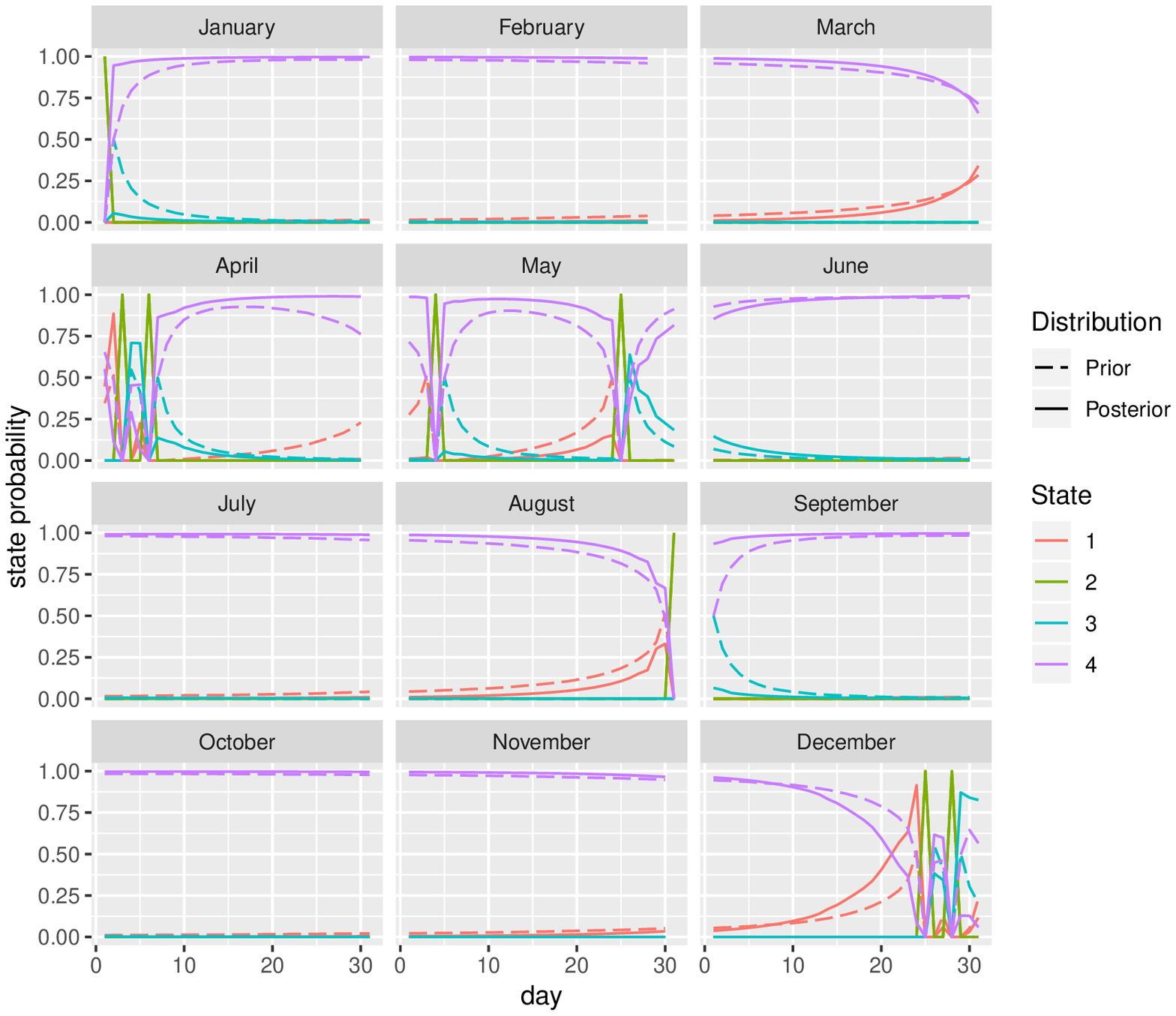}}
\caption{\label{fig:s_post_2015_LB3}Prior and posterior for the state sequence $S_t$ for load band 3 in an illustrative year (2015).}
\end{figure}

Clearly the pointwise posterior mode for the state sequence cannot give any indication of the uncertainty in the state allocation. Therefore, using load band 3 as an example, Figure~\ref{fig:s_post_2015_LB3} shows the posterior distribution for the states $S_t$ in a representative year (2015), with the prior distribution overlaid. There is a marked difference between the prior and posterior in the period around a public holiday. Although the prior treats all public holidays as if they are (near) exchangeable, there is strong evidence of differences between public holidays in the posterior. This shows the advantage of allowing the extent of proximity days to be unknown rather than periods of common, fixed duration around each public holiday. The corresponding plots for load bands 1 and 2 are given in Figures~\ref{fig:s_post_2015_LB1} and \ref{fig:s_post_2015_LB2} of the Supplementary Materials. Whilst the latter shows similar patterns to load band 3, there is very little difference between the prior and posterior for load band 1, though the prior does assign slightly more mass to the pre- and post-holiday states in the neighbourhood of public holidays. This suggests that there is little evidence about the presence, or otherwise, of a proximity effect in the data for load band 1.

\subsubsection{\label{subsubsec:pred_inf}Posterior predictive inference}
Consider a simplified version of the model described in Section~\ref{sec:model} which ignores the proximity effect and omits the pre- and post-holiday states (states 1 and 3) so that each day is classified either as a public holiday (state 2) or not (state 4). As discussed in Section~\ref{sec:intro}, many models from the literature are structured in this way, allowing for public holidays but not any protracted effect on neighbouring days. To illustrate the benefit of incorporating the proximity effect, we use the framework of posterior predictive checks \citep[][]{GCSDVR13} to compare inferences under our four-state NHMM with those computed under the simplified two-state model. In this framework, the basic idea is to measure the extent to which a model captures some data summary of interest by comparing its posterior predictive distribution to the value that was observed. In our case the (approximate) posterior predictive distribution is computed numerically based on an MCMC sample from the posterior of the model parameters by simulating replicated data sets in one-to-one correspondence with the posterior draws. If the model is able to capture adequately the aspect of the data summary of interest, the observed value will look plausible under its posterior predictive distribution.

\begin{figure}[t]
\centering
\scalebox{0.7}{\includegraphics{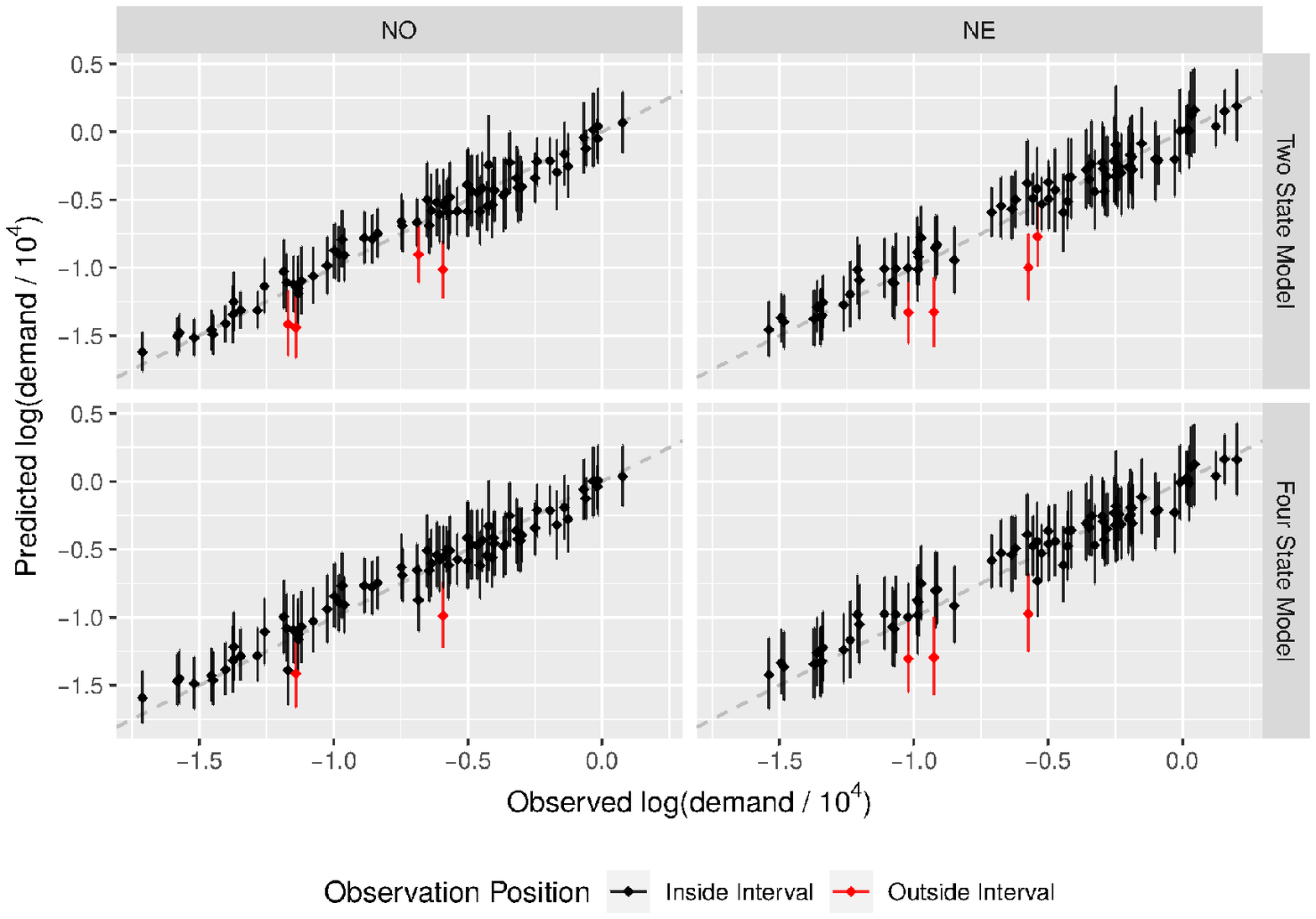}}
\caption{\label{fig:ypred_LB3_gap0}For load band 3 and each LDZ, posterior predictive means versus observed log gas demand for each public holiday in the observation period. Vertical bars extend to the 2.5\% and 97.5\% points in the posterior predictive distributions. Colours indicate whether the observation lay inside or outside the central 95\% of the posterior predictive distribution. Upper panels refer to the simple two-state model, lower panels on the four-state NHMM.}
\end{figure}

On public holidays and days which are not in the neighbourhood of a public holiday, both the two-state and four-state models seem to provide a good fit to the data for all load bands. For instance, in load band 3, for all the days during the observation period which were public holidays, Figure~\ref{fig:ypred_LB3_gap0} compares the posterior predictive distribution for log gas demand under each model to the values that were observed. Similarly, Figure~\ref{fig:ypred_LB3_gap10} in the Supplementary Materials provides an analogous comparison for days which were ten days away from a public holiday. In these plots, good model fit is indicated by the majority of points lying close to the unit diagonal, suggesting that most observations fall within the main body of their posterior predictive distribution. This is broadly true for both models in each LDZ, with only 1.15--6.67\% of observations lying outside the central 95\% of their associated posterior predictive distribution. 

\begin{figure}[t]
\centering
\scalebox{0.7}{\includegraphics{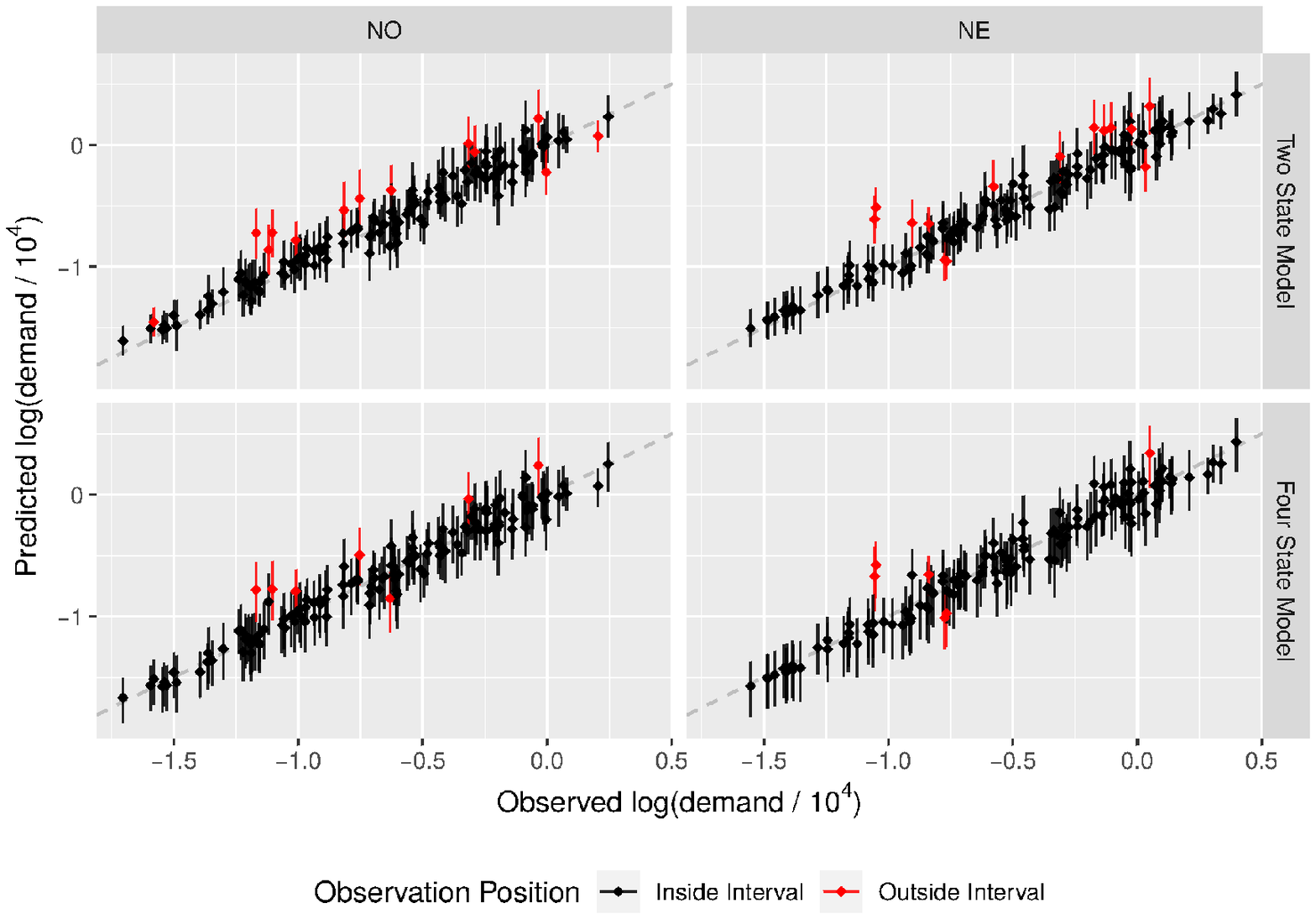}}
\caption{\label{fig:ypred_LB3_gap1}For load band 3 and each LDZ, posterior predictive means versus observed log gas demand for each day in the observation period which was one day from a public holiday. Vertical bars extend to the 2.5\% and 97.5\% points in the posterior predictive distributions. Colours indicate whether the observation lay inside or outside the central 95\% of the posterior predictive distribution. Upper panels are based on the simple two-state model, lower panels on the four-state NHMM.}
\end{figure}

The advantage of allowing a proximity effect becomes clear when we consider days close to a public holiday. For example, Figure~\ref{fig:ypred_LB3_gap1} shows the corresponding posterior predictive checks for all the days which were one day from a public holiday in load band 3. For each LDZ, it appears that the two-state model systematically overestimates the demand for gas, with 9.63\% and 10.67\% of observations lying outside (and mostly below) the central 95\% of their posterior predictive distributions in the NO and NE LDZs, respectively. In contrast, the posterior predictive densities under the four-state model generally support smaller values and are more diffuse, reflecting the additional uncertainty in the demand for gas that typically accompanies the period around public holidays. Correspondingly, the observations appear more plausible under the posterior predictive distributions, with only 5.19\% and 4.44\% lying outside the central 95\% of their posterior predictive distributions in the two LDZs. A similar effect, albeit less marked, is observed for days which are two or three days from a public holiday. Very similar conclusions can be drawn based on the analysis of the data from load band 2, plots for which are provided in Figures \ref{fig:ypred_LB2_gap0}--\ref{fig:ypred_LB2_gap1} of the Supplementary Materials. Not surprisingly, there is no notable difference between the two- and four-state models for load band 1, where our analysis suggested little evidence of a proximity effect. Supplementary Table~\ref{tab:pp} contains a summary of the results described in this section for all load bands.

\section{\label{sec:discuss}Discussion}
The energy sector in the UK is changing to harness greener technologies in the face of growing economic, environmental, societal and public health concerns. As a comparatively low cost source of energy, natural gas has an important role to play in this changing energy mix. Yet, clearly, its contribution is contingent on efficient operation of the gas industry. One of the ways in which operational efficiency and decision-making, more generally, can be improved is through improved forecasting of the demand for gas.

We have presented a novel model for daily gas demand which allows for a protracted effect of public holidays which extends into neighbouring days. A key feature of the model is that we allow the existence, duration and location of the proximity days to be unknown. This is achieved by modelling the data using a four-state NHMM, with cyclic dynamics, and states that represent pre- and post-holiday days, as well as (observable) public holidays and ``normal'' days. The dates of public holidays are always known in advance and we allow for this by making our model non-homogeneous, with transition probabilities that depend on the number of days to the next, and since the previous, public holiday. We establish an interpretable parameterisation for the logit transition probabilities and illustrate how this can be used to assign a prior distribution for the states that represents a flexible compromise between a model which does not allow for a proximity effect and one which fixes the dates over which a proximity effect is felt.

Conditional on the states, we model the natural logarithm of gas demand over two large geographical regions using a first-order vector autoregression, which is conditionally stationary. A novel feature of this model is the assumption of a symmetric, though non-diagonal, autoregressive coefficient matrix which greatly simplifies the geometry of the stationarity region, allowing it to be expressed as a unit square.

We applied the model to data from NGN, which is responsible for the distribution of gas to most of the North East of England, Northern Cumbria and Yorkshire. Assessment of model fit using posterior predictive checks highlighted the improvement gained by allowing for a proximity effect amongst commercial and small industrial customers. Similarly, the posterior distribution for the states revealed different patterns in the identification of proximity days between public holidays and across different groups of customers. This highlights the advantage of allowing uncertainty in the identification of proximity days compared to an inflexible approach that treats them as periods of common, fixed duration around each public holiday. The results of a preliminary version of the model are already being used by NGN in their annual medium-term forecasting exercise, with plans to extend the approach to consider the demand for gas by larger industrial customers.

\section*{Acknowledgements}
The authors thank NGN for providing the data and for many helpful discussions and comments during the research project. We also extend our gratitude to Matt Linsley, Manager of the Industrial Statistics Research Unit (ISRU) at Newcastle University, for his important role in organising and facilitating meetings with NGN.

\bibliographystyle{chicago}
\bibliography{refs}

\end{document}